\documentclass[twocolumn,prc,showpacs,preprintnumbers,superscriptaddress]{revtex4-1}

\usepackage{amsmath} 
\usepackage{amssymb}
\usepackage{amsfonts}
\usepackage{mathrsfs}
\usepackage{units}
\usepackage{multirow}
\usepackage{color}

\usepackage{dcolumn}
\usepackage{bm}
\usepackage{rotating} 

\makeatletter
\def\overbracket#1{\mathop{\vbox{\ialign{##\crcr\noalign{\kern3\p@}
    \downbracketfill\crcr\noalign{\kern3\p@\nointerlineskip}
    $\hfil\displaystyle{#1}\hfil$\crcr}}}\limits}
\def\downbracketfill{$\m@th
    \makesm@sh{\llap{\vrule\@height.7\p@\@depth2.3\p@\@width.7\p@}}%
    \leaders\vrule\@height.7\p@\hfill
    \makesm@sh{\rlap{\vrule\@height.7\p@\@depth2.3\p@\@width.7\p@}}$}
\makeatother
\newcommand{\disregard}[1]{}

\newcommand{\ohalf}{\tfrac{1}{2}}
\newcommand{\thalf}{\tfrac{3}{2}}
\newcommand{\fhalf}{\tfrac{5}{2}}
\newcommand{\shalf}{\tfrac{7}{2}}

\newsavebox{\tmpstrikebox}
\newlength{\tmpstrikelen}

\begin{document}

\title{No-core configuration-interaction model for the isospin-
and angular-momentum-projected states}

\author{W. Satu{\l}a}
\affiliation{Institute of Theoretical Physics, Faculty of Physics, University of Warsaw, ul. Pasteura 5,
PL-02-093 Warsaw, Poland}
\affiliation{Helsinki Institute of Physics, P.O. Box 64, FI-00014 University of Helsinki, Finland}

\author{P. B{\c a}czyk}
\affiliation{Institute of Theoretical Physics, Faculty of Physics, University of Warsaw, ul. Pasteura 5,
PL-02-093 Warsaw, Poland}

\author{J. Dobaczewski}
\affiliation{Institute of Theoretical Physics, Faculty of Physics, University of Warsaw, ul. Pasteura 5,
PL-02-093 Warsaw, Poland}
\affiliation{Helsinki Institute of Physics, P.O. Box 64, FI-00014 University of Helsinki, Finland}
\affiliation{Department of Physics, University of York, Heslington, York YO10 5DD, United Kingdom}
\affiliation{Department of Physics, P.O. Box 35 (YFL), University of Jyv\"askyl\"a, FI-40014  Jyv\"askyl\"a, Finland}

\author{M. Konieczka}
\affiliation{Institute of Theoretical Physics, Faculty of Physics, University of Warsaw, ul. Pasteura 5,
PL-02-093 Warsaw, Poland}

\date{\today}

\begin{abstract}
\begin{description}
\item[Background]
Single-reference density functional theory is very
successful in reproducing bulk nuclear properties like binding
energies, radii, or quadrupole moments throughout the entire periodic
table. Its extension to the multi-reference  level allows for
restoring symmetries and, in turn, for calculating transition rates.

\item[Purpose]
We propose a new no-core-configuration-interaction (NCCI) model
treating properly isospin and rotational symmetries. The model is
applicable to any nucleus irrespective of its mass and neutron- and
proton-number parity. It properly includes polarization effects caused
by an interplay between the long- and short-range forces acting in the
atomic nucleus.

\item[Methods]
The method is based on solving the Hill-Wheeler-Griffin equation
within a model space built of linearly-dependent states having good
angular momentum and properly treated isobaric spin. The states are generated by
means of the isospin and angular-momentum projection applied to a set
of low-lying (multi)particle-(multi)hole deformed Slater determinants
calculated using the self-consistent Skyrme-Hartree-Fock approach.

\item[Results]
The theory is applied to calculate energy spectra in
$N\approx Z$ nuclei that are relevant from the point of view of a
study of superallowed Fermi $\beta$-decays. In particular, a new set
of the isospin-symmetry-breaking corrections to these decays is
given.

\item[Conclusions]
It is demonstrated that the NCCI model is capable to capture main
features of low-lying energy spectra in light and medium-mass nuclei
using relatively small model space and without any local readjustment
of its low-energy coupling constants. Its flexibility and a range of
applicability makes it an interesting alternative to the conventional
nuclear shell model.

\end{description}
\end{abstract}

\pacs{
21.10.Hw, 
21.60.Jz, 
21.30.Fe, 
23.40.Hc, 
24.80.+y  
}
\maketitle

\section{Introduction}\label{intro}

Atomic nucleus is a self-bound finite system composed of neutrons and
protons that interact by means of short-range, predominantly
isospin-symmetry-conserving strong force and long-range
isospin-symmetry-breaking Coulomb force. In studies of phenomena
related to the isospin-symmetry violation in nuclei, capturing a
delicate balance between these two forces is of utmost importance.
This is particularly true when evaluating the
isospin-symmetry-breaking (ISB) corrections to superallowed
$\beta$-decays between isobaric analogue states,
[$I=0^+,T=1]\longrightarrow [I=0^+,T=1]$.

Such $\beta$-decays currently offer the most precise data that give
estimates of the vector coupling constant $G_V$ and leading element
$V_{ud}$ of the Cabibbo-Kobayashi-Maskawa (CKM) flavor-mixing
matrix~\cite{(Cab63),(Kob73)}. The uncertainty of $V_{ud}$ extracted
from the superallowed $\beta$-decays is almost an order of magnitude
smaller than that from neutron or pion decays~\cite{(Oli14)}. To test
the weak-interaction flavor-mixing sector of the Standard Model of
elementary particles, such precision is critical, because it allows
us to verify the unitarity of the CKM matrix, violation of which may
signal {\it new physics\/} beyond the Standard Model, see
Ref.~\cite{(Tow10a)} and references cited therein.

The isospin impurity of the nuclear wave function -- a measure of the
ISB -- is small. It varies from a fraction of a percent, in ground
states of even-even $N=Z$ light nuclei, to about six percent in the
heaviest known $N=Z$ system, $^{100}$Sn~\cite{(Sat09)}. Nevertheless,
its microscopic calculation poses a real challenge to theory. The
reason is that the isospin impurity originates from the long-range
Coulomb force that polarizes the entire nucleus and can be,
therefore, calculated only within so-called no-core approaches. In
medium and heavy nuclei, it narrows the possible microscopic models to
those rooted within the nuclear density functional theory
(DFT)~\cite{(Hoh64),(Koh65)}.

The absence of external binding requires that the nuclear DFT be
formulated in terms of intrinsic, and not laboratory densities. This,
in turn, leads to the spontaneous breaking of fundamental symmetries
of the nuclear Hamiltonian, including the rotational and isospin
symmetries, which in finite systems must be restored. Fully quantal
calculations of observables, such as matrix elements of
electromagnetic transitions or $\beta$-decay rates, require symmetry
restoration. In most of practical applications, this is performed
with the aid of the generalized Wick's theorem~\cite{(Bla86)}. Its
use, however, leads to the energy density functionals (EDFs) being
expressed in terms of the so-called transition densities, that is, to
a multi-reference (MR) DFT. Unfortunately, the resulting MR EDFs are,
in general, singular and require regularization, which still lacks
satisfactory and practical solution, see, e.g.,
Refs.~\cite{(Lac09),(Ben09),(Sat14b)}. An alternative way of building
a non-singular MR theory, the one that we use in the present work,
relies on employing the EDFs derived from a {\it true\/} interaction,
which then acquires a role of the EDF generator~\cite{(Dob15)}. The
results presented here were obtained using in this role the
density-independent Skyrme interaction SV~\cite{(Bei75)}, augmented
by the tensor terms (SV$_T$)~\cite{(Sat14b)}.

Over the last few years we have developed the MR DFT approach based
on the angular-momentum and/or isospin projections of single Slater
determinants. The model, below referred to as {\it static\/}, was
specifically designed to treat rigorously the conserved rotational
symmetry and, at the same time, tackle the explicit Coulomb-force
mixing of good-isospin states. These unique approach allowed us to
determine the isospin impurities in $N\approx Z$
nuclei~\cite{(Sat09)} and ISB corrections to superallowed
$\beta$-decay matrix elements~\cite{(Sat11),(Sat12)}.

In this paper, following upon preliminary results announced at
several conferences~\cite{(Sat13a),(Sat14),(Sat14d)}, we introduce a
next-generation {\it dynamic\/} variant of the approach, which we
call no-core configuration-interaction (NCCI) model. It constitutes a
natural extension of the static MR DFT model, and allows for mixing
states that are projected from different self-consistent Slater
determinants. The model is an analogue of the
generator-coordinate-method mixing of symmetry-projected states, see,
e.g., Ref.~\cite{(Bal14b)}, however, it rather addresses the mixing
of states representing low-lying (multi)particle-(multi)hole
excitations.

There are several cases when, to perform reliable calculations, such
dynamic approach is indispensable. One of the most important ones
relates to different possible shape-current orientations, which
within the static variant of the model appear in odd-odd
nuclei~\cite{(Sat12)}. The configuration mixing is also needed
to resolve the issue of unphysical ISB corrections to the analogous
states of the $A=38$ isospin triplet~\cite{(Sat11),(Sat12)}.

The states that are mixed have good angular momenta and, at the same
time, include properly evaluated Coulomb isospin mixing; hence, the
extended model treats hadronic and Coulomb interactions on the same
footing. The model is based on a truncation scheme dictated by the
self-consistent deformed Hartree-Fock (HF) solutions, and can be used
to calculate spectra, transitions, and $\beta$-decay rates in any
nucleus, irrespective of its even or odd neutron and proton numbers.

We begin by giving in Sec.~\ref{sec:ncci} a short overview of the
theoretical framework of our NCCI model. In Sec.~\ref{sec:isb}, a new
set of the ISB corrections to the canonical set of superallowed
$\beta$-decay is presented. As compared to our previous
results~\cite{(Sat12)}, the new set includes mixing of reference
states corresponding to different shape-current orientations in
odd-odd $N=Z$ nuclei. In Sec.~\ref{sec:spec}, applications involving
mixing of several low-energy (multi)particle-(multi)hole excitations
are discussed. Here, we determined low-spin energy spectra in
selected nuclei relevant to high-precision tests of the
weak-interaction flavor-mixing sector of the Standard Model. The
calculations were performed for: $^{6}$Li and $^{8}$Li nuclei
(Sec.~\ref{ssec:Li}), $A$=38 Ar, K, and Ca nuclei
(Sec.~\ref{ssec:A38}), $^{42}$Sc and $^{42}$Ca nuclei
(Sec.~\ref{ssec:A42}), and $^{62}$Ga and $^{62}$Zn nuclei
(Sec.~\ref{ssec:A62}. Summary and perspectives are given in
Sec.~\ref{sec:sum}.

\section{The no-core con\-fi\-gu\-ra\-tion-in\-te\-rac\-tion model}\label{sec:ncci}

In an even-even nucleus, the Slater determinant representing the
ground-state is uniquely defined. However, in an odd-odd nucleus, the
conventional mean-field (MF) theory, which builds upon separate
Slater determinants for neutrons and protons, faces problems. First,
there is no Slater determinant representing the $T=1, I=0^+$ state in
odd-odd $N=Z$ nuclei, see~\cite{(Sat12),(Sat13c)}. In our approach, this obstacle is
removed by projecting good isospin from the so-called antialigned
Slater determinant. This configuration has, by construction, no net
alignment and manifestly breaks the isospin symmetry, being an almost
fifty-fifty mixture of the $T=0$ and $T=1$ states. By projecting the
isospin, the needed $T=1$ component is thus recovered. Second,
the antialigned states are not uniquely defined. In
the general case of a triaxial nucleus, there exist three
linearly-dependent Slater determinants, built of valence neutron and
proton single-particle (s.p.) states that individually carry angular momenta
aligned along the long, intermediate, or short axes of the core. In
our calculations, no tilted-axis antialigned solutions were ever
found so far.

In the static approach, the only way to cope with this ambiguity is
to calculate three independent $\beta$-decay matrix elements and to
take the average of the resulting $\delta_{\rm C}$
values~\cite{(Sat12)}. Such a solution is not only somewhat
artificial, but also increases the theoretical uncertainty of the
calculated ISB corrections. This drawback has motivated our
development of the dynamic model, which allowed for the mixing of
states projected from the three reference states
$\varphi_k$, with angular momenta aligned along the
Cartesian $k=X$, $Y$, and $Z$ axes of the intrinsic reference frame.
In the dynamic model, the mixing matrix elements between these states
are derived from the same Hamiltonian that is used to calculate them.
Then, the model further evolved towards a full NCCI model, in which
we allow for a mixing of states projected from arbitrary low-lying
(multi)particle-(multi)hole Slater determinants $\varphi_i$.
This final variant has all features of the {\it no-core\/} shell
model, in that it uses the two-body effective interaction, but it also
includes a full non-perturbative Coulomb force and its
basis-truncation scheme is dictated by the self-consistent deformed
HF solutions.

\begin{figure}[htb]
\centering
\includegraphics[width=0.8\columnwidth]{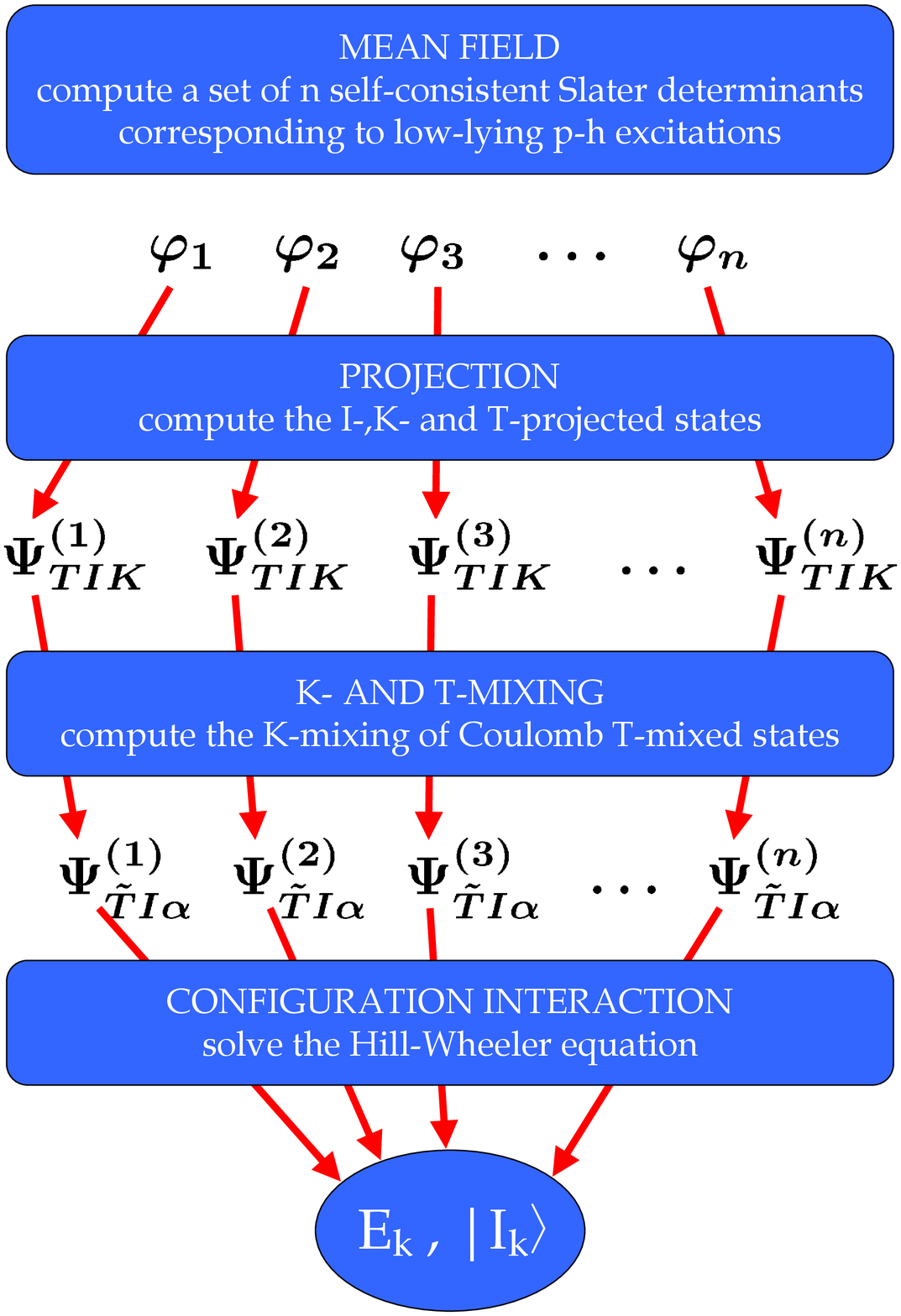}
\caption{(Color online) Computational scheme of the NCCI model. See text for
details.}
\label{fig01}
\end{figure}

The computational scheme of the NCCI model is sketched in
Fig.~\ref{fig01}. It proceeds in four major steps:
\begin{itemize}

\item
First, a set of relevant low-lying (multi)particle-(multi)hole HF
states $\{ \varphi_i \}$ is calculated along with their HF energies
$e^{{\rm (HF)}}_i$. States $\{ \varphi_i \}$ form a subspace of
reference states for subsequent projections.

\item
Second, the  projection techniques are applied to the set of
reference states $\{ \varphi_i \}$, so as to determine the family of
states $\{\Psi_{TIK}^{(i)}\}$ having good isospin $T$, angular
momentum $I$, and angular-momentum projection on the intrinsic axis
$K$.

\item
Third, states $\{\Psi_{TIK}^{(i)}\}$ are mixed, so as to properly
take into account the $K$ mixing and Coulomb isospin mixing -- this
gives the set of good angular-momentum states $\{\Psi_{\tilde{T}I\alpha}^{(i)}\}$ of
the static model~\cite{(Sat11),(Sat12)}. Here we label them with the
dominating values of the isospin, $\tilde{T}$, and auxiliary quantum
numbers $\alpha$.

\item
Finally, the results of the dynamic model correspond to mixing
non-orthogonal states $\{\Psi_{\tilde{T}I\alpha}^{(i)}\}$ for all
configurations $i$, and for all values of $\tilde{T}$ and $\alpha$.
This is performed by solving the Hill-Wheeler equation
${\cal{H}}|I_k\rangle=E_k{\cal{N}}|I_k\rangle$~\cite{(Rin80)} in the
{\it collective space\/} spanned by the {\it natural states\/}
corresponding to sufficiently large eigenvalues of the norm matrix
${\cal{N}}$. This is the same technique that is used in the code to handle
the $K$-mixing alone. The method is described in details in
Ref.~\cite{(Dob09d)}.

\end{itemize}
We note here that all wave functions considered above correspond to
good neutron ($N$), and proton ($Z$) numbers, and thus to a good
third component of the isospin, $T_z=\ohalf(N-Z)$. We also note that
the configuration interaction, which is taken into account in the
last step, could have also equivalently been performed by directly
mixing the projected states $\{\Psi_{TIK}^{(i)}\}$. The procedure
outlined above simply aims to obtain separately the results of the
static and dynamic model. The NCCI calculations discussed below were
performed using a new version of the HFODD solver~\cite{(Sat16)},
which was equipped with the NCCI module. This new implementation was
based on the previous versions of the
code~\cite{(Dob09d),(Sch12),(Sch14)}.

Numerical stability of the method depends on necessary truncations
of the model space. In this work, numerically unstable solutions are
removed by truncating the natural states corresponding to
small eigenvalues of the norm matrix $N$. It means that only the
natural states corresponding to the eigenvalues of the norm matrix
that are larger than certain externally provided cut-off parameter
$\zeta$ are used to built the so-called collective space.
Although such a truncation procedure gives reliable values of the
energy, a full stability of the method still requires further
studies. Other methods, e.g., based on truncating high-energy
states $\{\Psi_{\tilde{T}I\alpha}^{(i)}\}$, or combined methods
involving both truncations simultaneously, need to be studied as
well.

\section{A new set of the ISB corrections to superallowed $\beta$-decays}\label{sec:isb}

In this section we present results obtained within the NCCI model,
which pertain to removing the uncertainty related to ambiguities in
the shape-current orientation. Similar to our previous applications
within the static model, the ground-states (g.s.) of even-even
nuclei, $| I=0, T\approx 1, T_z = \pm 1 \rangle$, are approximated by
the Coulomb $T$-mixed states,
\begin{eqnarray}
  |I=0, T\approx 1, T_z = \pm 1 \rangle
  &=&                          \Psi_{\tilde{T}=1,I=0,K=0}^{(1)}
\nonumber \\
  &=&   \sum_{T\geq 1} c^{(1)}_{T} \Psi_{T,I=0,K=0}^{(1)},
\end{eqnarray}
which were angular-momentum projected from the MF g.s.\ $\varphi_1$
of the even-even nuclei, obtained in the self-consistent Hartree-Fock
(HF) calculations. States $\varphi_1$ are always unambiguously
defined by filling in the pairwise doubly degenerate levels of
protons and neutrons up to the Fermi level. In the calculations, the
Coulomb $T$-mixing was included up to $T=4$.

Within our dynamic model, the corresponding isobaric analogues in
$N=Z$ odd-odd nuclei, $|I=0, T\approx 1, T_z = 0 \rangle$, were
approximated by
\begin{eqnarray}\label{oddphi}
  | I=0, T\approx 1, T_z = 0  \rangle
  &=& \!\!\!\! \sum_{k=X,Y,Z} \sum_{\tilde{T}=0,1,2} c^{(k)}_{\tilde{T}}
   \Psi_{\tilde{T},I=0,K=0}^{(k)}.
 \nonumber \\
\end{eqnarray}
The underlying MF states $\varphi_k$ were taken as the
self-consistent Slater determinants $|\bar \nu \otimes \pi ; k
\rangle$ (or  $| \nu \otimes \bar \pi ; k \rangle$) representing the
antialigned configurations corresponding to different shape-current
orientations $k=X,Y,Z$. Let us recall that the antialigned states
are constructed by placing the odd neutron and odd proton in the
lowest available time-reversed (or signature-reversed)
s.p.\ orbits. These states are manifestly breaking
the isospin symmetry. Using them is the only way to reach the $| I=0,
T\approx 1, T_z = 0  \rangle$ states in odd-odd $N=Z$ nuclei. The
reason is that, within a conventional MF approach with separate
proton and neutron Slater determinants, these states are not
representable by single Slater determinants, see discussion in
Ref.~\cite{(Sat13c)}.

For odd-odd nuclei, mixing coefficients $c^{(k)}_{\tilde{T}}$ in
Eq.~(\ref{oddphi}) were determined by solving the Hill-Wheeler
equation. In the mixing calculations, we only included states
$\Psi_{\tilde{T},I=0,K=0}^{(k)}$ with dominating isospins of
$\tilde{T}=0,1$ and 2, that is, the Hill-Wheeler equation was solved
in the space of six or nine states for axial and triaxial states,
respectively. We recall that each of states
$\Psi_{\tilde{T},I=0,K=0}^{(k)}$ contains all Coulomb-mixed good-$T$
components $\Psi_{{T},I=0,K=0}^{(k)}$.

\begin{figure}[htb]
\centering
\includegraphics[width=0.8\columnwidth]{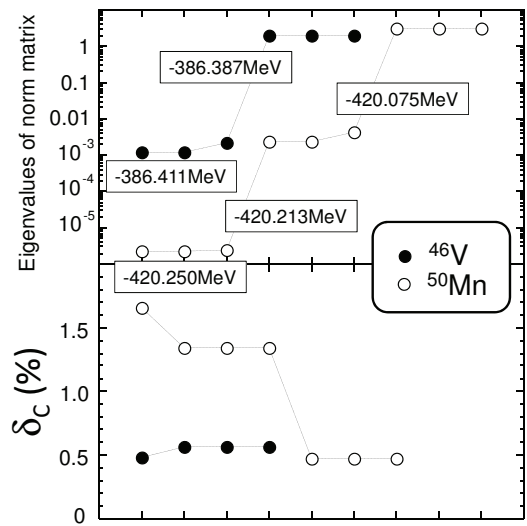}
\caption{Upper part shows eigenvalues of the norm matrix obtained in the NCCI calculations for
the $I=0$ states of odd-odd nuclei. Depicted are typical results obtained for
two representative examples of axial ($^{46}$V) and triaxial ($^{50}$Mn)
nuclei. The boxes give values of eigenenergies obtained by including
three, six, or nine eigenvalues of the norm matrix. Lower part shows a dependence of
the ISB corrections
to superallowed $^{46}$V$\rightarrow ^{46}$Ti and $^{50}$Mn$\rightarrow ^{50}$Cr decays on a number
of collective states retained in the mixing calculations. The number of states retained in the mixing
decreases from the left to the right hand side.
}
\label{fig:norms}
\end{figure}

The three states corresponding to a given dominating isospin are
linearly dependent. One may therefore argue that the physical
subspace of the $I=0$ states should be three dimensional. In the
calculations, all six or nine eigenvalues of the norm matrix ${\cal{N}}$ are
nonzero, but the linear dependence of the reference states is clearly
reflected in the pattern they form. For two representative examples
of axial ($^{46}$V) and triaxial ($^{50}$Mn) nuclei, this is depicted
in Fig.~\ref{fig:norms}. Note, that the eigenvalues group into two or
three sets, each consisting of three similar eigenvalues.
Note also, that the differences between the sets are large, reaching
three-four orders of magnitude.
Lower part of the figure illustrates dependence of the calculated
ISB corrections, $\delta_{\rm C}$, on a number of the collective states
retained in the mixing. As shown, the calculated corrections are becoming stable
within a subspace consisting five (or less) highest-norm
states. Following this result, we have decided to retain in the
mixing calculations only three collective states built upon the three
eigenvectors of the norm matrix corresponding to the largest
eigenvalues.

Based on this methodology, we calculated the set of the superallowed
transitions, which are collected in Tables~\ref{tab:delta1} and
\ref{tab:delta2}. Table~\ref{tab:delta1} shows the empirical
$ft$ values, calculated ISB corrections, and so-called
nucleus-independent reduced lifetimes,
\begin{equation}\label{ftnew}
   {\cal F}t \equiv ft(1+\delta_{\rm R}^\prime)(1+\delta_{\rm NS} -\delta_{\rm C})
  = \frac{K}{2 G_{\rm V}^2 (1 + \Delta^{\rm V}_{\rm R})},
\end{equation}
where $\delta_{\rm R}^\prime$ and $\delta_{\rm NS}$ are the radiative
corrections~\cite{(Tow08)}. Errors of ${\cal F}t$ include errors of
the empirical $ft$ values~\cite{(Tow10),(Har14)}, radiative
corrections $\delta_{\rm R}^\prime$ and $\delta_{\rm
NS}$~\cite{(Tow08)}, and our uncertainties estimated for the
calculated values of $\delta_{\rm C}$.

Except for transitions $^{14}$O$\rightarrow$$^{14}$N  and
$^{42}$Sc$\rightarrow$$^{42}$Ca, all ISB corrections were
calculated using the prescription sketched above. For the decay
of a spherical nucleus $^{14}$O, the reference state is uniquely
defined and thus the mixing of orientations was not necessary, whereas
for that of $^{42}$Sc, an ambiguity of choosing its reference state
is not related to the shape-current orientation. For both cases, the
values and errors of $\delta_{\rm C}$ were taken from
Ref.~\cite{(Sat12)}. For the remaining cases, to account for
uncertainties related to the basis size and collective-space cut-off,
we assumed an error of 15\%. This is larger than the 10\%
uncertainties related to the basis size only, which were assumed
in Ref.~\cite{(Sat12)}.

Systematic errors related to the form and parametrization of the
functional itself were not included in the error budget. Moreover,
similarly to our previous works~\cite{(Sat11),(Sat12)}, transition
$^{38}$K$\rightarrow$$^{38}$Ar was disregarded. We recall that for
this transition, the calculated value of the ISB correction is
unacceptably large because of a strong mixing of Nilsson levels
originating from the $d_{3/2}$ and $s_{1/2}$ sub-shells.  The problem
can be partially cured by performing configuration-interaction
calculations, see Ref.~\cite{(Sat14d)} and discussion in
Sect.~\ref{ssec:A38}.

\begin{table*}
\caption[A]{\label{tab:delta1} Results of calculations performed for nuclei, for which the
superallowed transitions have been measured.  Listed are
empirical $ft$ values~\protect\cite{(Har14)}; calculated ISB
corrections $\delta_{\rm C}^{{\rm (SV)}}$ and the corresponding
${\cal F}t$ values; empirical corrections $\delta_{\rm C}^{{\rm (exp)}}$
calculated using Eq.~(\protect\ref{expdelt});
contributions coming from the individual transitions to the $\chi^2$
budget in the confidence-level test.
As in Ref.~\protect\cite{(Har14)}, we give two digits of the calculated errors
of the ${\cal F}t$ values.}
\begin{ruledtabular}
\begin{tabular}{lrlrrrlr}
Parent    &  \multicolumn{1}{c}{$ft$} & $\quad$ &
$\delta_{\rm C}^{{\rm (SV)}}~\strut$  &  ${\cal F}t~~~~~\strut$  &  $\quad$ & \multicolumn{1}{c}{$\delta_{\rm C}^{{\rm (exp)}}~\strut$}  &  $\chi^2_i$ \\
nucleus   &   \multicolumn{1}{c}{(s)} &         &
(\%)$~~~\strut$                       &       (s)$~~~~~\strut$   &          &   \multicolumn{1}{c}{(\%)$~~~\strut$}                      &              \\
\hline
$T_z=-1:$ &                        &&            &             &&            &      \\
$^{10}$C  &  3042(4)               &&   0.579(87)& 3064.5(52)  && 0.37(15)   &  3.5 \\
$^{14}$O  &  3042.3(27)            &&   0.303(30)& 3072.3(33)  && 0.36(6)    &  0.0 \\
$^{22}$Mg &  3052(7)               &&   0.270(41)& 3081.4(72)  && 0.62(23)   &  1.4 \\
$^{34}$Ar &  3053(8)               &&    0.87(13)& 3063.6(91)  && 0.63(27)   &  1.3 \\[3pt]
$T_z=0: $ &                        &&            &             &&            &      \\
$^{26}$Al &  3036.9(9)             &&   0.329(49)& 3071.8(20)  && 0.37(4)    &  0.8 \\
$^{34}$Cl &  3049.4(12)            &&    0.75(11)& 3067.6(38)  && 0.65(5)    & 10.9 \\
$^{42}$Sc &  3047.6(14)            &&    0.77(27)& 3069.2(85)  && 0.72(6)    &  3.1 \\
$^{46}$V  &  3049.5(9)             &&   0.563(84)& 3075.1(32)  && 0.71(6)    &  1.3 \\
$^{50}$Mn &  3048.4(12)            &&   0.476(71)& 3076.5(32)  && 0.67(7)    &  2.4 \\
$^{54}$Co &  3050.8($^{+11}_{-15}$)&&   0.586(88)& 3075.6(36)  && 0.75(8)    &  1.3 \\
$^{62}$Ga &  3074.1(15)            &&    0.78(12)& 3093.1(48)  && 1.51(9)    & 43.2 \\
$^{74}$Rb &  3085(8)               &&    1.63(24)&   3078(12)  && 1.86(27)   &  0.3 \\[3pt]
          &             &&
              $\overline{{\cal F}t}=$ &  3073.7(11) &&$\chi^2 =$  & 69.5 \\
          &             &&
                      $|V_{\rm ud}|=$ & 0.97396(25) &&$\chi_d^2 =$&  6.3 \\
          &             &&
                                      & 0.99937(65) &&            &
\end{tabular}
 \end{ruledtabular}
\end{table*}

\begin{table}
\caption[A]{\label{tab:delta2}Similar as in
Table~\protect\ref{tab:delta1} but for the transitions that are
either unmeasured or measured with insufficient accuracy to be used
for the SM tests.}
\begin{ruledtabular}
\begin{tabular}{lcccrcrc}
Parent    & $\delta_{\rm C}^{\rm (SV)}$  & $\quad$ & Parent    & $\delta_{\rm C}^{\rm (SV)}$ \\
nucleus   &                              &         & nucleus   &                             \\
          &       (\%)                   &         &           &       (\%)                  \\
\hline
$T_z=-1:$ &          & & $T_z=0: $ &           \\
$^{18}$Ne &  1.37(21)& & $^{18}$F  &  1.22(18) \\
$^{26}$Si & 0.427(64)& & $^{22}$Na & 0.335(50) \\[3pt]
$^{30}$S  &  1.24(19)& & $^{30}$P  &  0.98(15)
\end{tabular}
 \end{ruledtabular}
\end{table}

To conform with the analyzes of Hardy and Towner (HT) and Particle Data
Group, the average value $\overline{{\cal F}t} = 3073.7(11)$s was
calculated using the Gaussian-distribution-weighted formula. This
leads to the value of $|V_{\rm ud}| = 0.97396(25)$, which is in a very good agreement
both with the Hardy and Towner result~\cite{(Har14)},  $|V_{\rm
ud}^{{\rm (HT)}}| = 0.97425(22)$, and central value obtained from
the neutron decay $|V_{\rm ud}^{(\nu )}| =
0.9746(19)$~\cite{(Nak10)}. By combining the value of $|V_{\rm ud}|$
calculated here with those of $|V_{\rm us}| = 0.2253(8)$ and $|V_{\rm
ub}| = 0.00413(49)$ of the 2014 Particle Data Group~\cite{(Oli14)},
one obtains
\begin{equation}\label{ckm}
       |V_{\rm ud}|^2 +  |V_{\rm us}|^2  + |V_{\rm ub}|^2 =  0.99937(65),
\end{equation}
which implies that the unitarity of the first row of the  CKM matrix
is satisfied with a precision better than 0.1\%. Note that, in spite
of differences between individual values of $\delta_{\rm C}$, the
values of $\overline{{\cal F}t}$ and $|V_{\rm ud}|$ obtained here are
in excellent agreement with the results of our previous
works~\cite{(Sat11),(Sat12)}.

The last two columns of Table~\ref{tab:delta1} show results of the
confidence-level (CL) test, as proposed in  Ref.~\cite{(Tow10)}. The CL
test is based on the assumption that the CVC hypothesis is valid up
to at least $\pm 0.03$\%, which implies that a set of
structure-dependent corrections should produce statistically
consistent set of ${\cal F}t$-values. Assuming the validity of the
calculated corrections  $\delta_{\rm NS}$~\cite{(Tow94)}, the
empirical ISB corrections can be defined as:
\begin{equation}\label{expdelt}
\delta_{\rm C}^{{\rm (exp)}}  = 1 + \delta_{\rm NS}
- \frac{\overline{{\cal F}t}}{ft(1+\delta_{\rm R}^\prime)}.
\end{equation}
By the least-square minimization of the appropriate $\chi^2$, and
treating the value of $\overline{{\cal F}t}$ as a single adjustable
parameter, one can attempt to bring the set of empirical values
$\delta_{\rm C}^{{\rm (exp)}}$ as close as possible to the set of
$\delta_{\rm C}$.

The  empirical ISB corrections deduced in this way are tabulated in
Table~\ref{tab:delta1}. The table also lists  individual
contributions to the $\chi^2$ budget, whereas the total $\chi^2$ per
degree of freedom ($\chi_d^2=\chi^2/n_d$ for $n_d=11$) is $\chi_d^2 = 6.3$. This number is
considerably smaller than the number quoted in our previous
work~\cite{(Sat12)}, but much bigger than those obtained within (i) perturbative-model
reported in Ref.~\cite{(Tow10)}  (1.5), (ii)
shell model with the Woods-Saxon radial wave
functions  (0.4) \cite{(Tow08)}, (iii) shell model with
Hartree-Fock radial wave functions  (2.0)
\cite{(Orm96),(Har09)}, (iv) Skyrme-Hartree-Fock with RPA (2.1)
\cite{(Sag96a)}, and relativistic Hartree-Fock plus RPA model
(1.7) \cite{(Lia09)}. It is
worth stressing that, as before, our value of $\chi^2 /n_d$ is
deteriorated by two transitions that strongly violate the CVC
hypothesis, $^{62}$Ga$\rightarrow ^{62}$As and
$^{34}$Cl$\rightarrow ^{34}$S. These transitions give the 62\% and
15\% contributions to the total error budget, respectively. Without them,
we would have obtained $\chi_d^2 = \chi^2 /9 = 1.7$.

\section{Low-energy spectra of selected nuclei}\label{sec:spec}

In this section, we present a short overview of results obtained
using the NCCI approach. Since the model is based on simultaneous
isospin and angular-momentum projections, it is particularly well
suited to study $N\approx Z$ nuclei. These nuclei are of paramount
importance for stringent many-body tests of the weak sector of the
Standard Model~\cite{(Har14),(Nav09a)}. Besides, they show specific
structural features, like the Wigner energy or Nolen-Schiffer
anomaly, which are difficult to reproduce within state-of-the-art
nuclear models, in particular those rooted in a standard DFT.

A major goal of this work is to pin down strong and weak points of
the NCCI approach proposed here. Hence, instead of performing a
detailed study of a single nucleus, with many configurations being
mixed, we decided to use a modest number of configurations and apply
the model to a rather broad set of nuclei, starting from very light
systems like $^{6,8}$Li up to $^{62}$Zn. By adding additional
configurations, the present results can certainly be refined. We
believe, however, that such refinements will not affect the physical
conclusions drown in this work.

To efficiently track the MF configurations and to improve convergence
properties of self-consistent calculations, all reference states used
in the NCCI calculations below were determined assuming the
conservation of parity and signature symmetries. For the $A\leq
42$($A=62$) nuclei, we employed the s.p.\ basis consisting of
$N=10(12)$ spherical harmonic oscillator shells, respectively.

\subsection{Lithium isotopes: $^{6}$Li and $^{8}$Li}\label{ssec:Li}

The Slater determinants, which we selected for the NCCI calculations
in these two very light nuclei, are listed in Table~\ref{tab:Li}. For
the sake of simplicity, the states are labeled by spherical quantum
numbers $p_{1/2}$ or $p_{3/2}$ that dominate in the s.p.\ wave
functions of the odd-proton and odd-neutron states. It turns out that
such a labeling constitutes an intuitive and relatively unambiguous
way to describe the configurations, even in cases of large
deformations where the Nilsson picture formally prevails. The
strategy behind selecting the reference configurations is to cover
basic combinations of neutron or proton particle-hole (p-h)
excitations having all possible alignments predicted by a simple
$K$-scheme.

\begin{table}[tbh]
\caption{Properties of the reference Slater determinants in $^{6}$Li and
$^{8}$Li, numbered by index $i$ and labeled by spherical
quantum numbers of particle states above $^4$He.
Listed are: the HF energies E$_{\rm HF}$ in MeV,
quadrupole deformations $\beta_2$, triaxiality parameters $\gamma$,
and neutron and proton s.p.\ alignments, $j_\nu$ and $j_\pi$,
together with their orientations $k$ in the intrinsic frame.}
\label{tab:Li}
\renewcommand{\arraystretch}{1.3}
\begin{tabular}{ccccrrrr}
\hline
$i$ & $|^{6}$Li$;i\rangle$        &    E$_{\rm HF}$  &   $\beta_2$   &  $\gamma$    & $j_\nu$ & $j_\pi$ & $k$ \\
\hline
 1  &  $\nu p_{3/2}\otimes \pi p_{3/2}$  &$-$25.972   &   0.008       &     0$^\circ$   &     $-$0.50 & 1.50 & Z\\
 2  &  $\nu p_{3/2}\otimes \pi p_{3/2}$  &$-$26.787   &   0.330       &     0$^\circ$   &        0.50 & 0.50 & Z\\
 3  &  $\nu p_{3/2}\otimes \pi p_{3/2}$  &$-$26.510   &   0.216       &    60$^\circ$   &     $-$1.50 & 1.50 & Y\\
 4  &  $\nu p_{3/2}\otimes \pi p_{3/2}$  &$-$27.244   &   0.207       &    60$^\circ$   &        1.50 & 1.50 & Y\\
 5  &  $\nu p_{3/2}\otimes \pi p_{3/2}$  &$-$26.846   &   0.090       &    60$^\circ$   &        1.50 & 0.50 & Y\\
\hline
$i$ & $|^{8}$Li$;i\rangle$        &     E$_{\rm HF}$  &   $\beta_2$   &  $\gamma$     & $j_\nu$ & $j_\pi$ & $k$   \\
\hline
 1  &  $\nu p_{3/2}\otimes \pi p_{3/2}$   &$-$39.081  &   0.381       &     0$^\circ$    &$-$1.50 & 0.50 & Z\\
 2  &  $\nu p_{1/2}\otimes \pi p_{3/2}$   &$-$34.041  &   0.361       &     0$^\circ$    &   0.50 & 0.50 & Z\\
 3  &  $\nu p_{3/2}\otimes \pi p_{3/2}$   &$-$39.025  &   0.356       &     0$^\circ$    &   1.50 & 0.50 & Z\\
 4  &  $\nu p_{3/2}\otimes \pi p_{3/2}$   &$-$35.680  &   0.027       &     0$^\circ$    &$-$1.50 & 1.50 & Z\\
 5  &  $\nu p_{1/2}\otimes \pi p_{3/2}$   &$-$33.443  &   0.352       &     0$^\circ$    &$-$0.50 & 0.50 & Z\\
\hline
\hline
\end{tabular}
\end{table}

\begin{figure}[htb]
\centering
\includegraphics[width=0.8\columnwidth]{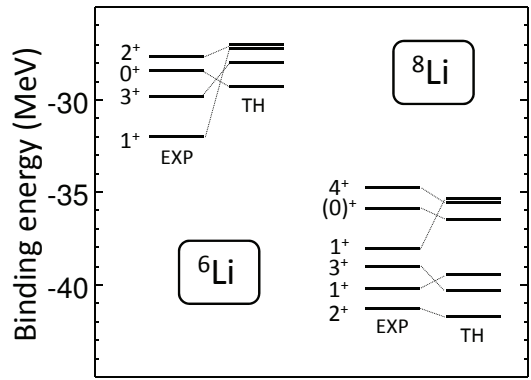}
\caption{Comparison between experimental and theoretical energy
spectra of $^6$Li and $^8$Li.}
\label{fig:Li}
\end{figure}

Results of our calculations are shown in Fig.~\ref{fig:Li}. In the
case of $^{6}$Li, theory clearly disagrees with data, with respect to
both the ordering and values of energies. Let us first discuss the
$T=0$ multiplet, composed of the $1^+$ and $3^+$ states. The ground
state of $^6$Li has quantum numbers $I=1^+, T=0$ and the experimental
total energy of this state is $-31.995$\,MeV. In calculations, the
lowest $I=1^+$ state is placed above the lowest $I=0^+,T=1$ and
$I=3^+,T=0$ solutions, and its energy of $-27.037$\,MeV is almost
$5$\,MeV higher than in experiment. For comparison, the calculated
energy of the $I=3^+,T=0$ member of the isoscalar multiplet is only
$1.8$\,MeV higher than in experiment. Hence, it is quite evident that
the model lacks the isoscalar pairing $I=1,T=0$ correlations,
cf.~Ref.~\cite{(Cor13)}. In the $N=Z$ nuclei, the model, or the
underlying mean-field, seems to favor the maximally aligned $T=0$
configurations. In Sec.~\ref{ssec:A42} we demonstrate that the
results obtained for $^{42}$Sc corroborate these conclusions.

It is worth recalling here that in the context of searching for
possible fingerprints of collective isoscalar $pn$-pairing phase in
$N\approx Z$ nuclei, the isoscalar pairing, or deuteron-like
correlations, were intensely discussed in the literature,
see~Refs.~\cite{(Goo72),(Eng96),(Sat97a),(Sat01a),(Sat01b)} and
references cited therein. In particular, the isoscalar $pn$-pairing
was considered to be the source of an additional binding energy that
could offer a microscopic explanation of the so-called Wigner
energy~\cite{(Sat97)} -- an extra binding energy along the $N=Z$
line, which is absent in the self-consistent MF mass models. In spite of
numerous recent works following these early developments attempting to
explain the isoscalar $pn$-pairing correlations
and the Wigner energy, see Refs.~\cite{(San12),(Ben13c),(Car14),(Ben14),
(Sam15a),(Del15)} and refs. cited therein, the problem still lacks a
satisfactory solution.

There are at least two major reasons for that: (i) an incompleteness
of the HFB (HF) approaches used so far, which consider the $pn$
mixing only in the particle-particle channel, see discussion in
Ref.~\cite{(Sat13c)}, and (ii) difficulties in evaluating the role of
beyond-mean-field correlations. Recently, within the RPA including
$pn$ correlations, the latter problem was addressed in
Ref.~\cite{(Car14)}. Their systematic study of the isoscalar and
isovector multiplets in magic and semi-magic nuclei rather clearly
indicated a missing relatively strong $T=0$ pairing. This seems to be
in line with our NCCI model findings concerning description of
$T=0,I=1$ states, but seems to contradict the conclusions of
Ref.~\cite{(Ben13c),(Ben14)}.

Concerning the $T=1$ multiplet consisting of the $0^+$ and $2^+$
states, the theory tends to overbind the $0^+$ state by $0.8$\,MeV and
underbind the $2^+$ state by $0.4$\,MeV.  This level of agreement
is much better than the one obtained for the
isoscalar multiplet. It should be rated as fair, but not fully
satisfactory. It is, therefore, interesting and quite surprising to see
that the addition of two neutrons in $^8$Li seems to change the situation quite
radically. Indeed, in this nucleus, for both the binding energies and distribution of
levels below 5\,MeV, the overall agreement between
theory and experiment is very
satisfactory, even if the calculated
$1_1^+$ and $3_1^+$ states are interchanged,
see Fig.~\ref{fig:Li}. The largest disagreement is obtained for the
$1_2^+$ state, where the theory underbinds experiment by almost
3\,MeV. The states $0_1^+$, $2_2^+$, and  $4_1^+$ are predicted at
the excitation energies of 5.3, 4.7, and 6.2\,MeV, respectively, in
fair agreement with the data.

\subsection{$0^+$ states in $^{38}$Ca, $^{38}$K, and
$^{38}$Ar}\label{ssec:A38}

Recently, Park {\it et al.\/}~\cite{(Par14)} performed high-precision
measurement of the superallowed $0^+\longrightarrow 0^+$ Fermi decay
of $^{38}$Ca$\rightarrow$$^{38}$K, see also~\cite{(Bla14)}. The
reported $ft$ value of 3062.3(68)\,s was measured with a relative
precision of $\pm$0.2\%, which is sufficient for testing and
determining the parameters of electroweak sector of the Standard
Model. This piece of data is the first, after almost a decade,
addition to a set of canonical $0^+\longrightarrow 0^+$ Fermi
transitions, which are used to determine $|V_{ud}|$. Moreover, being
a mirror partner to superallowed $0^+\longrightarrow 0^+$ Fermi
transition $^{38}$K$\rightarrow$$^{38}$Ar, it allows for sensitive
tests of the ISB corrections and, in turn, for assessing quality of
nuclear models used to compute the ISBs~\cite{(Par14)}.

Unfortunately, using the DFT with the SV Skyrme functional, which
gives a strong mixing between the $2s_{1/2}$ and $1d_{3/2}$ orbits,
it is difficult to determine the ISB corrections to the
$^{38}$K$\rightarrow$$^{38}$Ar and $^{38}$Ca$\rightarrow$$^{38}$K
superallowed transitions. In particular, in our previous static DFT
calculations, the ISB corrections turned out to be of the order of
9\%, and thus were disregarded~\cite{(Sat11),(Sat12)}.

In Ref.~\cite{(Sat14d)}, we presented preliminary results of the
NCCI study of $^{38}$Ca and $^{38}$K. Here we extend them to
calculations that include three low-lying antialigned reference
configurations in $^{38}$K and four configurations in both $^{38}$Ca
and $^{38}$Ar. Basic properties of these reference states are listed
in Table~\ref{tab:A38}.

\begin{table}[tbh]
\caption{Similar as in Table~\protect\ref{tab:Li}, but for $^{38}$K,
$^{38}$Ca, and $^{38}$Ar. Here, the reference Slater determinants are
labeled by the Nilsson quantum numbers pertaining to dominant
components of the hole states below $^{40}$Ca. The first excited
state in $^{38}$Ar, marked by asterisk, was converged with a weak
quadrupole constraint. }
\label{tab:A38}
\renewcommand{\arraystretch}{1.3}
\begin{tabular}{cclcrrrr}
\hline
$i$ & $|^{38}$K$;i\rangle$        & $\Delta$E$_{\rm HF}$  &   $\beta_2$        &  $\gamma$       &  $j_\nu$ & $j_\pi$ & $k$ \\
\hline
 1  &  $|202\thalf\rangle^{-2}$     &       0.000         &      0.083       &    60$^\circ$   &     $-$0.50 &    0.50 & Y\\
 2  &  $|220\ohalf\rangle^{-2}$     &       1.380         &      0.035       &     0$^\circ$   &        0.50 & $-$0.50 & Z\\
 3  &  $|211\ohalf\rangle^{-2}$     &       1.559         &      0.042       &     0$^\circ$   &     $-$1.50 &    1.50 & Z\\
\hline
$i$ & $|^{38}$Ca$;i\rangle$        & $\Delta$E$_{\rm HF}$  &   $\beta_2$     &  $\gamma$    &  $j_\nu$ & $j_\pi$ & $k$  \\
\hline
 1  &  $|200\ohalf\rangle^{-2}$    &       0.000      &        0.088         &     60$^\circ$  &  0 & 0 & --    \\
 2  &  $|200\ohalf\rangle^{-2}$    &       0.762      &        0.006         &      0$^\circ$  &  0 & 0 & --    \\
 3  &  $|211\ohalf\rangle^{-2}$    &       1.669      &        0.045         &      0$^\circ$  &  0 & 0 & --    \\
 4  & $|220\ohalf\rangle^{-1}\otimes|202\thalf\rangle^{-1}$
                                   &       2.903      &        0.015         &     60$^\circ$  &  0 & 0 & --    \\
\hline
$i$ & $|^{38}$Ar$;i\rangle$        & $\Delta$E$_{\rm HF}$   &   $\beta_2$    &  $\gamma$   &  $j_\nu$ & $j_\pi$ & $k$  \\
\hline
 1  &  $|200\ohalf\rangle^{-2}$    &       0.000      &        0.088         &     60$^\circ$  &  0 & 0 & --    \\
 2  &  $|200\ohalf\rangle^{-2}$    &     0.651$^{(*)}$&        0.002         &     46$^\circ$  &  0 & 0 & --    \\
 3  &  $|211\ohalf\rangle^{-2}$    &       1.600      &        0.045         &      0$^\circ$  &  0 & 0 & --    \\
 4  & $|220\ohalf\rangle^{-1}\otimes|202\thalf\rangle^{-1}$
                                   &       2.754      &        0.017         &     60$^\circ$  &  0 & 0 & --    \\
\hline
\hline
\end{tabular}
\end{table}

\begin{figure}[htb]
\centering
\includegraphics[width=0.8\columnwidth]{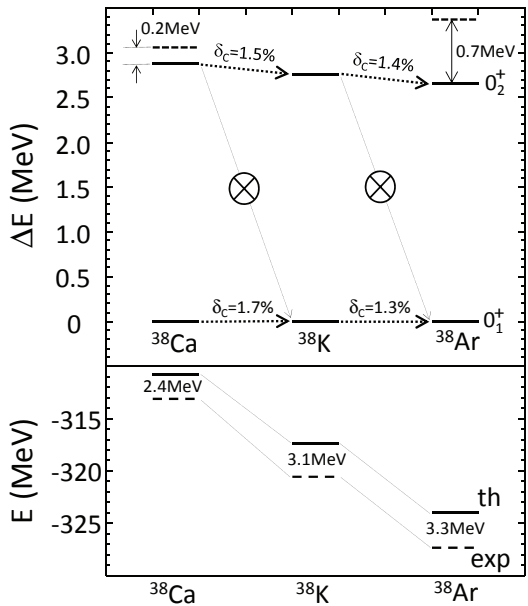}
\caption{The top panel shows excitation energies of the $0_2^+$
states with respect to the $0_1^+$ states in the $A=38$ isobaric
triplet nuclei: Ca, K, and Ar. Theoretical predictions and
data~\protect\cite{(ensdf)} are shown with solid and dashed lines,
respectively. Calculated values of $\delta_{\rm C}$ are also shown.
Decays $0_2^+ \rightarrow 0_1^+$ indicated in the figure are
predicted to be strongly hindered. The bottom panel shows a similar
comparison between the total binding energies of the $0_1^+$ states ($N=10$
harmonic oscillator shells were used).}
\label{fig:A38}
\end{figure}

Results of our NCCI calculations, including the binding energies of the
lowest $0_1^+$ states, excitation energies of the first excited
$0_2^+$ states, and the ISB corrections to superallowed
$\beta$-decays, are visualized in Fig.~\ref{fig:A38}. The total
binding energies of the $0_1^+$ states in these three nuclei are
underestimated by circa 1\%. Concerning the first excited $0_2^+$
states, our model works very well in $^{38}$Ca. In this nucleus, the
measured excitation energy, $\Delta E_{\rm EXP}=3057(18)$\,keV, is
only 186\,keV larger than the calculated one, $\Delta E_{\rm
TH}=2871$\,keV. Note, however, that the calculated excitation
energies of the $0_2^+$ states are predicted to decrease with increasing
$T_z$, at variance with the data. In turn, the difference between
experimental and theoretical excitation energies of the $0_2^+$ in
$^{38}$Ar grows to approximately 0.7\,MeV.

The ISB corrections $\delta_{\rm C}$ to the
$^{38}$Ca$\rightarrow$$^{38}$K transitions between the
$0_1^+\rightarrow 0_1^+$ and $0_2^+\rightarrow 0_2^+$ states are
equal to 1.7\% and 1.5\%, respectively. As compared to our previous
static model, which for the $0_1^+$ states was giving an unacceptably
large correction of 8.9\%, the NCCI result is strongly reduced.
Nevertheless, it is still almost twice larger than that of Towner and
Hardy~\cite{(Tow08)}, who quote the value of 0.77(7)\%.

Similar results were obtained for the $^{38}$K$\rightarrow$$^{38}$Ar
transitions, where the calculated corrections are 1.3\%
($0_1^+\rightarrow 0_1^+$) and 1.4\% ($0_2^+\rightarrow 0_2^+$).
Again, as compared to the static variant of our model, the value for
the $0_1^+\rightarrow 0_1^+$ transition is strongly reduced, but it
is considerably larger than the Towner and Hardy result of 0.66(6)\%.
Nevertheless, we see that the NCCI model removes, at lest partially, pathologies
encountered in the static variant.

\subsection{A=42 nuclei: $^{42}$Sc and $^{42}$Ca}\label{ssec:A42}

Within the conventional shell model, the $^{42}$Ca and $^{42}$Sc nuclei
are treated as two-body systems above the core of $^{40}$Ca. Hence,
they are often used by the shell-model community to adjust the
isoscalar $T$=0;\,$I$=1, 3, 5, 7, and isovector $T$=1;\,$I$=0, 2, 4, 6
matrix elements within the $f_{7/2}$ shell. Here, we use these
nuclei to test our NCCI model but, at least at this stage, without an
intention of refitting the interaction. The aim of this exercise is to
capture global trends and tendencies, which may allow us to identify
systematic features of the NCCI model in describing these
seemingly simple nuclei. From the perspective of our approach, such
tests are by no means trivial, because these nuclei are here treated within
the full core-polarization effects included, cf.~discussion in
Refs.~\cite{(Zal08),(Tar14)}.

\begin{figure}[htb]
\centering
\includegraphics[width=0.8\columnwidth]{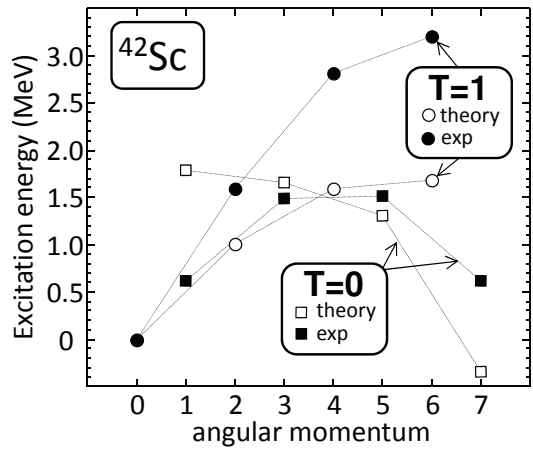}
\caption{Excitation energies of the isovector (circles)
and isoscalar (squares) multiplets in $^{42}$Sc with respect to the
$0^+$ state. Theoretical and experimental results are marked with open
and filled symbols, respectively.}
\label{fig:42sc}
\end{figure}

\begin{figure}[htb]
\centering
\includegraphics[width=0.8\columnwidth]{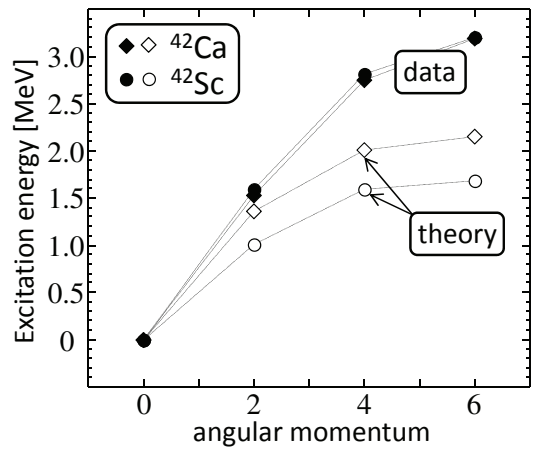}
\caption{Same as in Fig.~\protect\ref{fig:42sc}, but for the isovector multiplets
in $^{42}$Sc (circles) and $^{42}$Ca (diamonds).}
\label{fig:A42}
\end{figure}

The results of the NCCI calculations for the isovector and isoscalar
multiplets in $A=42$ nuclei are depicted in Figs.~\ref{fig:42sc}
and~\ref{fig:A42}, and collected in Table~\ref{tab:A42}. The
reference states used in the calculation for $^{42}$Sc are listed in
Table~\ref{tab:42sc}. They cover all fully aligned ($K_\nu =K_\pi$)
states, which are almost purely isoscalar, all possible
antialigned states ($K_\nu = - K_\pi$), and two $K=1$ aligned states.
The antialigned states manifestly violate the isospin symmetry and,
as discussed in Ref.~\cite{(Sat10)}, are approximately fifty-fifty
mixtures of the isoscalar and isovector components. The $K=1$ aligned
states also violate the isospin symmetry.

\begin{table}[tbh]
\caption{Excitation energies (total energies) of low-lying states
(ground states) in $^{42}$Sc and $^{42}$Ca. For $^{42}$Sc, we show
calculated and experimental energies of the isovector
($I$=0$^+$, 2$^+$, 4$^+$, and 6$^+$) and isoscalar
($I$=1$^+$, 3$^+$, 5$^+$, and 7$^+$) multiplets. For $^{42}$Ca, we show
the analogous energies of the isovector multiplet. All energies are
in MeV.}
\label{tab:A42}
\renewcommand{\arraystretch}{1.3}
\centering\begin{tabular}{rrrrrr}
\hline
              & $^{42}$Sc                      & $^{42}$Sc
              & $^{42}$Ca                      & $^{42}$Ca                       \\
$I^{\rm \pi}$ & $\Delta$E$_{\rm I}^{\rm (th)}$ & $\Delta$E$_{\rm I}^{\rm (exp)}$
              & $\Delta$E$_{\rm I}^{\rm (th)}$ & $\Delta$E$_{\rm I}^{\rm (exp)}$ \\
\hline
 0$^+$        &$-$352.961    &$-$354.687       &$-$360.200       &  $-$361.895                 \\
 2$^+$        &   1.012      &    1.586        &    1.357        &       1.525                 \\
 4$^+$        &   1.590      &    2.815        &    2.005        &       2.752                 \\
 6$^+$        &   1.696      &   (3.200)       &    2.154        &       3.189                 \\
\hline
 1$^+$        &   1.785      &    0.611        &                  &                            \\
 3$^+$        &   1.656      &    1.490        &                  &                            \\
 5$^+$        &   1.336      &    1.510        &                  &                            \\
 7$^+$        &$-$0.347      &    0.617        &                  &                             \\
\hline
\hline
\end{tabular}
\end{table}

The following three general conclusions can be drown from the results
presented in Fig.~\ref{fig:42sc}:
\begin{itemize}
\item
The model lacks isoscalar pairing $T=0,I=1$ correlations. A similar
deficiency was already observed in $^{6}$Li, Sec.~\ref{ssec:Li}.

\item
The model strongly prefers fully aligned isoscalar $T=0,I_{\rm
max}=7$ states. Again, the conclusion is consistent with the one
drown from the calculated spectrum of $^{6}$Li.

\item
The energy range spanned by the isovector states, $\Delta E_{T=1} =
E_{T=1,I=6} - E_{T=1,I=0}$, is by a factor of two smaller in theory
than in experiment. It is not obvious, however, whether this
difference signalizes that the model underestimates the isovector
pairing correlations, overbinds the stretched (aligned)
configurations, or both.

\end{itemize}

\begin{table}[tbh]
\caption{Similar as in Table~\protect\ref{tab:Li}, but for $^{42}$Sc.
Here, the reference Slater determinants correspond to configurations
$\nu f_{7/2}\otimes \pi f_{7/2}$, and are labeled by intrinsic $K$
quantum numbers of valence neutrons and protons as $|\nu;
K_\nu\rangle \otimes |\pi ; K_\pi\rangle$. Reference states $i$=1--4
correspond to antialigned configurations, $K_\nu = - K_\pi$, thus
carrying no net intrinsic alignment. Reference states $i$=5--8
represent aligned configurations, $K_\nu = K_\pi$, thus having the total
alignments of 7, 5, 3, and 1, respectively. The remaining two
configurations $i$=9--10 carry net alignments of 1. The Table also
lists the HF energies $\Delta$E$_{\rm I=|K|}$ relative to the $|\nu
;\ohalf \rangle \otimes |\overline{ \pi ;\ohalf }\rangle$ solution.
The last column shows excitation energy of the lowest $I=|K|$ state
projected from a given Slater determinant.}
\label{tab:42sc}
\renewcommand{\arraystretch}{1.3}
\centering\begin{tabular}{rccccc}
\hline
$i$ &   $|^{42}$Sc$;i\rangle $        &$\Delta$E$_{\rm HF}$  &   $\beta_2$   &  $\gamma$    & $\Delta$E$_{\rm I=|K|}$ \\
\hline
 1  &  $|\nu ;\ohalf\rangle \otimes |\overline{\pi ;\ohalf}\rangle$
                &    0.000   &   0.063       &     0   &   0.000   \\
 2  &  $|\nu ;\thalf\rangle \otimes |\overline{\pi ;\thalf}\rangle$
                &    0.802   &   0.031       &     0   &   0.561   \\
 3  &  $|\nu ;\fhalf\rangle \otimes |\overline{\pi ;\fhalf}\rangle$
                &    0.986   &   0.008       &    60   &   0.551   \\
 4  &  $|\nu ;\shalf\rangle \otimes |\overline{\pi ;\shalf}\rangle$
                &    0.759   &   0.062       &    60    &  0.085   \\
\hline
 5  &  $|\nu ;\shalf\rangle \otimes |\pi ;\shalf\rangle$
                                        & $-$0.929   &   0.061       &    60    & -0.647   \\
 6  &  $|\nu ;\fhalf\rangle \otimes |\pi ;\fhalf\rangle$
                                        &    0.082   &   0.007       &    60    &  1.160   \\
 7  &  $|\nu ;\thalf\rangle \otimes |\pi ;\thalf\rangle$
                                        &    0.345   &   0.032       &     0    &  1.594   \\
 8  &  $|\nu ;\ohalf\rangle \otimes |\pi ;\ohalf\rangle$
                                        &    0.340   &   0.060       &     0    &  1.719   \\
\hline
 9  &  $|\nu ;\thalf\rangle \otimes |\pi ;-\ohalf\rangle$
                                        &    0.716   &   0.043       &     0    &  2.164   \\
10  &  $|\nu ;\fhalf\rangle \otimes |\pi ;-\thalf\rangle$
                                        &    0.986   &   0.011       &     0    &  2.338   \\
\hline
\hline
\end{tabular}
\end{table}

\begin{figure}[htb]
\centering
\includegraphics[width=0.8\columnwidth]{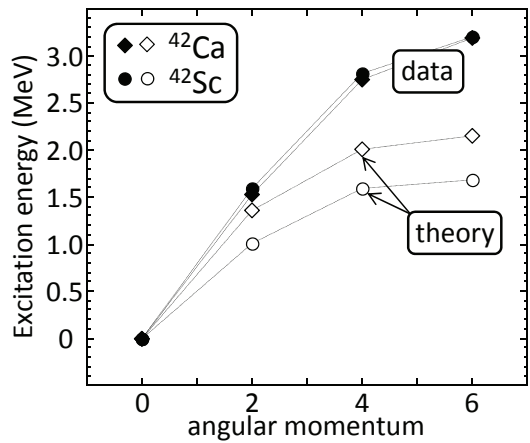}
\caption{Schematic illustration of the interplay between the
primary physical ingredients contributing to the excitation energy
of the intruder state within our model. See text for details.}
\label{fig:intruder}
\end{figure}

In the case of $^{42}$Ca, we focused on calculating the excitation
energies of the $0^+$ states, addressing, in particular, the question
of structure and excitation energy of the intruder
configuration. Experimentally, the intruder configuration is observed
at very low excitation energy of 1.843\,MeV, see Ref.~\cite{(Had13)}
and references cited therein. In the calculations presented below we
assumed that the structure of intruder state is associated with
(multi)particle-(multi)hole excitations across the $N=Z=20$ magic gap,
which in $^{40}$Ca is of the order of 7.0\,MeV, see Ref.~\cite{(Zal08)}
and references cited therein. The mechanism bringing the intruder
configuration down in energy is sketched in
Fig.~\ref{fig:intruder}.

The energy needed to elevate particles from the $d_{3/2}$ subshell to
$f_{7/2}$ is at (near)spherical shape reduced by the energy
associated with the spontaneous breaking of spherical symmetry in the
intruder configuration, and further, by a rotational correction
energy associated with the symmetry restoration. Owing to the
configuration interaction, an additional gain in energy is expected
too. The rotational correction and configuration interaction are also
expected to lower slightly the MF g.s.\ energy. As shown in
Fig.~\ref{fig:intruder}, the final value of the intruder excitation
energy is an effect of rather a delicate interplay of several
factors. Therefore, it is not surprising that the intruder states
pose a real challenge for both the state-of-the-art nuclear shell
models and MF-rooted theories.

In the NCCI calculations presented below, we mix states projected
from the antialigned configurations that are listed in
Table~\ref{tab:42ca}. The reference states can be divided into two
classes. The first four  configurations  do not involve any
cross-shell excitations. They correspond to the $|K\rangle \otimes
|\overline{K}\rangle$ 0p-0h $(\nu f_{7/2})^2$ configurations with
magnetic quantum number of $K$=1/2, 3/2, 5/2, and 7/2, respectively. The
three remaining configurations are the lowest  MF  configurations
involving two $f_{7/2}^4 d_{3/2}^{-2}:\, (\nu f_{7/2})^2\otimes (\pi
f_{7/2})^2 \otimes (\pi d_{3/2})^{-2} $, four $f_{7/2}^6
d_{3/2}^{-4}:\, (\nu f_{7/2})^4\otimes (\nu d_{3/2})^{-2} \otimes
(\pi f_{7/2})^2 \otimes (\pi d_{3/2})^{-2} $, and six $f_{7/2}^8
d_{3/2}^{-6}:\, (\nu f_{7/2})^4\otimes (\nu d_{3/2})^{-2} \otimes
(\pi f_{7/2})^4 \otimes (\pi d_{3/2})^{-4} $ holes in $d_{3/2}$
shell, respectively.

\begin{table}[tbh]
\caption{Similar as in Table~\protect\ref{tab:Li}, but for $^{42}$Ca.
Here, the Slater determinants are labeled by spherical
quantum numbers pertaining to active neutron orbitals. The last
column shows excitation energies of the lowest $0^+$ states projected
from a given Slater determinant.}
\label{tab:42ca}
\renewcommand{\arraystretch}{1.3}
\centering\begin{tabular}{cccccc}
\hline
$i$ & $|^{42}$Ca$;i\rangle$        &$\Delta$E$_{\rm HF}$  &   $\beta_2$   &  $\gamma$    & $\Delta$E$_{I=0}$ \\
\hline
 1  &     $|\ohalf\rangle \otimes |\overline{\ohalf}\rangle$   &    0.000   &   0.069       &     0$^\circ$   &   0.000   \\
 2  &     $|\thalf\rangle \otimes |\overline{\thalf}\rangle$   &    0.516   &   0.033       &     0$^\circ$   &   0.765   \\
 3  &     $|\fhalf\rangle \otimes |\overline{\fhalf}\rangle$   &    0.544   &   0.007       &    60$^\circ$   &   0.770   \\
 4  &     $|\shalf\rangle \otimes |\overline{\shalf}\rangle$   &    0.084   &   0.061       &    60$^\circ$    &  0.315   \\
\hline
 5  &  $f_{7/2}^4\, d_{3/2}^{-2}$
                                        &   10.001   &   0.288       &    14$^\circ$    &  6.860   \\
 6  &  $f_{7/2}^6\, d_{3/2}^{-4}$
                                        &   10.986   &   0.414       &    22$^\circ$    &  6.498   \\
 7  &  $f_{7/2}^8\, d_{3/2}^{-6}$
                                        &   14.937   &   0.542       &    12$^\circ$    &  9.619   \\
\hline
\hline
\end{tabular}
\end{table}

\begin{figure}[htb]
\centering
\includegraphics[width=0.8\columnwidth]{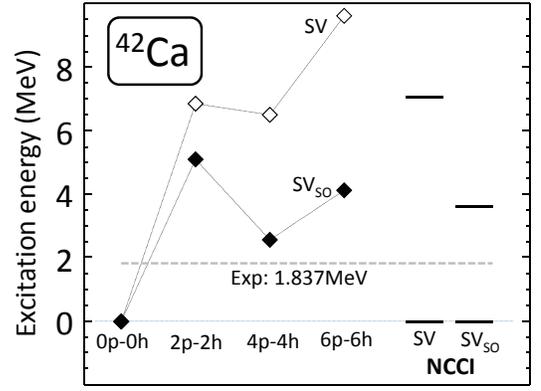}
\caption{The lowest $0^+$ states projected from $(f_{7/2})^2$ (0p-0h),
$(f_{7/2})^4 (d_{3/2})^{-2}$ (2p-2h), $(f_{7/2})^6 (d_{3/2})^{-4}$
(4p-4h), and  $(f_{7/2})^8 (d_{3/2})^{-6}$ (6p-6h) reference states.
Open (filled) diamonds refer to calculations performed using the SV and
SV$_{\rm SO}$ functionals, respectively. These results do not include
configuration mixing. The right part shows excitation energies of the
intruder state obtained within the NCCI theory with SV and  SV$_{\rm
SO}$ interactions.}
\label{fig:42ca}
\end{figure}

The results of the NCCI calculations in $^{42}$Ca are depicted in
Figs.~\ref{fig:A42} and~\ref{fig:42ca} and collected in
Tables~\ref{tab:A42} and \ref{tab:42ca}. Figure~\ref{fig:A42} shows the
$I=0^+,2^+,4^+$, and 6$^+$ states -- the isovector $T$=1 multiplet --
obtained within the NCCI calculations involving only $(\nu f_{7/2})^2$
reference states. The results are qualitatively similar to those in
$^{42}$Sc. In both cases, theoretical spectra are compressed as
compared to data. Detailed quantitative comparison reveals, however,
surprisingly large differences between the theoretical and experimental
spectra.

First, the energy differences $\delta E_{\rm I} = \Delta E_I (
^{42}{\rm Ca}) - \Delta E_I ( ^{42}{\rm Sc})$ for $I=2^+, 4^+, 6^+$
are positive (negative) in theory (experiment), respectively. Second,
the absolute values of $|\delta E_{\rm I}|$ are a few times larger in
theory as compared to the data. It means that the model tends to
overestimate the ISB effects in clearly an unphysical manner. This
influences the ISB correction to the $0^+\longrightarrow 0^+$ Fermi
$\beta$-decay matrix element, which in the present NCCI calculation
rises to $\delta_{\rm C} \approx 2.2$\%. Most likely, the unphysical
component in the ISB effect is related to the time-odd polarizations and
matrix elements originating from these fields, which are essentially
absent in even-even systems. One should also remember that in the
Skyrme functionals, including, of course, the SV force used here, the
time-odd terms are very purely constrained.

Figure~\ref{fig:42ca} shows the $0^+$ states calculated using
functionals SV. The left part of the figure depicts the
lowest $0^+$ states projected from the reference states $(f_{7/2})^2$
(0p-0h), $(f_{7/2})^4 (d_{3/2})^{-2}$ (2p-2h), $(f_{7/2})^6
(d_{3/2})^{-4}$ (4p-4h), and  $(f_{7/2})^8 (d_{3/2})^{-6}$ (6p-6h).
These results do not include configuration mixing. Note, that
symmetry restoration itself changes the optimal intruder
configuration to $(\nu f_{7/2})^4\otimes (\nu d_{3/2})^{-2} \otimes
(\pi f_{7/2})^2 \otimes (\pi d_{3/2})^{-2} $ as compared to MF, which
favors  $(\nu f_{7/2})^2\otimes (\pi f_{7/2})^2 \otimes (\pi
d_{3/2})^{-2} $.

The right part of the figure shows excitation energies of the
intruder states as obtained within the NCCI calculations. Here, all
reference states listed in Table~\ref{tab:42ca} were included. For the
SV force, the excitation energy of the lowest intruder configuration
equals 7.5\,MeV, and exceeds the data by 5.7\,MeV. The main reason of
the disagreement is related to an unphysically large $N=Z=20$ shell
gap: the bare $N=20$ gap deduced directly from the s.p.\ HF levels in
$^{40}$Ca equals as much as 11.5\,MeV. Its value exceeds the
experimental gap by almost 4.5\,MeV (for an overview of experimental
data, see Ref.~\cite{(Zal08)} and references cited therein). It is
therefore not surprising that the combined effects of deformation and
rotational correction are unable to compensate for the large energy
needed to lift the particles from the $d_{3/2}$ to $f_{7/2}$ shell,
see Fig.~\ref{fig:intruder}.

To investigate interplay between the s.p.\ and collective
effects, we repeated the NCCI calculations using the functional
SV$_{\rm SO}$, which differs from SV in a single aspect, namely, we
increased its spin-orbit strength by a factor of 1.2. This readjustment allows to reduce
a disagreement between theoretical and experimental binding energies in $N\approx Z$
$sd$ and lower-$pf$ shell nuclei to $\pm$1\% level as shown in Ref.~\cite{(Kon15)}.
When applied to the heaviest $N = Z$ nucleus $^{100}$Sn and its neighbor $^{100}$In it gives
827.710 MeV and 833.067 MeV what is in an impressive agreement with the experimental binding energies
equal 825.300 MeV (833.110 MeV) in $^{100}$Sn ($^{100}$In), respectively.
Ability to reproduce masses is among the most important indicators of a quality of DFT-based models.
Such a readjustment of the SO strength is also the simplest
and most efficient mechanism allowing us to reduce the magic $Z=N=20$
gap~\cite{(Zdu05)}. For the SV$_{\rm SO}$ force, the bare gap equals
9.6\,MeV, which is by almost 1.9\,MeV smaller than the original SV
gap, but still it is much larger, by circa 2.6\,MeV, than the
experimental value. Results of the NCCI calculations obtained using
functional SV$_{\rm SO}$ are shown in Fig.~\ref{fig:42ca}. Now the
projected and NCCI calculations both favor the configuration $(\nu
f_{7/2})^4\otimes (\nu d_{3/2})^{-2} \otimes (\pi f_{7/2})^2 \otimes
(\pi d_{3/2})^{-2} $. We also note that for both the SV and SV$_{\rm
SO}$ functionals, geometrical properties of the reference states
(deformations) are very similar.

When discussing the influence of various effects on the final
position of the intruder state, it is worth stressing the role of the
symmetry restoration. The rotational correction lowers the intruder
state by 4.9\,MeV, bringing its excitation energy to 2.3\,MeV, which
is only 0.5\,MeV above the experiment. However, after the
configuration mixing, the excitation energy of the intruder state
increases to about 3.6\,MeV, that is, it becomes again 1.7\,MeV
higher as compared to data. This is due to the configuration mixing
in the ground state, which lowers its energy by almost 1\,MeV,
whereas it leaves the position of the intruder state almost
unaffected. The reason for that is the fact that the $(\nu
f_{7/2})^2$ antialigned reference states (states 1--4 in
Tables~\ref{tab:42ca} and \ref{tab:SVSO}) are almost linearly
dependent and thus mix relatively
strongly. Conversely, at deformations corresponding to the intruder
configuration $(\nu f_{7/2})^4\otimes (\nu d_{3/2})^{-2} \otimes (\pi
f_{7/2})^2 \otimes (\pi d_{3/2})^{-2} $, the Nilsson scheme prevails.
Therefore, the intruder configurations become almost linearly
independent and appear to mix very weakly. The amount of the mixing
was tested by performing additional calculations of matrix elements
between the lowest $(\nu f_{7/2})^4\otimes (\nu d_{3/2})^{-2} \otimes
(\pi f_{7/2})^2 \otimes (\pi d_{3/2})^{-2} $ configuration and the
excited configurations involving the same number of $(d_{3/2})^{-4}$
holes. All these matrix elements turned out to be negligibly small.

\begin{table}[tbh]
\caption{
Same as in Table~\protect{\ref{tab:42ca}}, but for the functional
SV$_{\rm SO}$.}
\label{tab:SVSO}
\renewcommand{\arraystretch}{1.5}
\centering\begin{tabular}{cccccc}
\hline
$i$ & $|^{42}$Ca$;i\rangle$                      &$\Delta$E$_{\rm HF}$     &  $\beta_2$ &  $\gamma$  & $\Delta$E$_{I=0}$ \\
\hline
 1  &     $|\ohalf\rangle \otimes |\overline{\ohalf}\rangle$  &    0.000   &   0.064    &  0$^\circ$ &  0.000   \\
 2  &     $|\ohalf\rangle \otimes |\overline{\ohalf}\rangle$  &    0.517   &   0.032    &  0$^\circ$ &  0.679   \\
 3  &     $|\ohalf\rangle \otimes |\overline{\ohalf}\rangle$  &    0.496   &   0.007    & 60$^\circ$ &  0.676   \\
 4  &     $|\ohalf\rangle \otimes |\overline{\ohalf}\rangle$  &    0.006   &   0.061    & 60$^\circ$ &  0.200   \\
\hline
 5  &     $f_{7/2}^4\, d_{3/2}^{-2}$                          &    8.399   &   0.276    & 15$^\circ$ &  5.085   \\
 6  &     $f_{7/2}^6\, d_{3/2}^{-4}$                          &    7.377   &   0.402    & 22$^\circ$ &  2.548   \\
 7  &     $f_{7/2}^8\, d_{3/2}^{-6}$                          &    9.955   &   0.532    & 15$^\circ$ &  4.103   \\
\hline
\hline
\end{tabular}
\end{table}

\subsection{A=62 nuclei: $^{62}$Zn and $^{62}$Ga}\label{ssec:A62}

\begin{figure}[htb]
\centering
\includegraphics[width=\columnwidth]{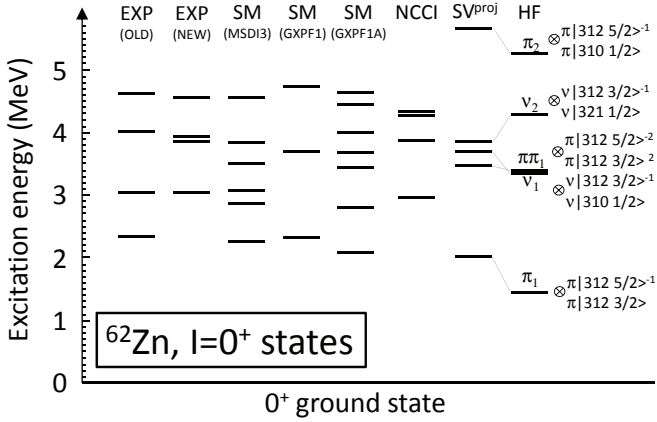}
\caption{The low-lying $0^+$ states in $^{62}$Zn.
The first two columns show old and new experimental data,
see~\cite{(Lea13)} for details. The next three columns collect the
results of the shell-model calculations using interactions MSDI3~\cite{(Koo77)},
GXPF1~\cite{(Hon02)}, and GXPF1A~\cite{(Hon04)},
respectively. The last three columns show results obtained within the NCCI approach,
angular-momentum projection, and pure HF method, respectively.
From Ref.~\cite{(Sat14d)}. }
\label{fig:62zn}
\end{figure}

For $^{62}$Zn, the results of the NCCI calculations of the low-lying
$0^+$ states were communicated in Ref.~\cite{(Sat14d)}. Here, for the
sake of completeness, we briefly summarize the
results obtained therein. The calculated spectrum of the $0^+$ states
below the excitation energy of 5\,MeV is shown in Fig.~\ref{fig:62zn}.
The NCCI calculations were based on six reference states that include:
the ground state, the two lowest neutron p-h excitations $\nu_1$ and
$\nu_2$, the two lowest proton p-h excitations $\pi_1$ and $\pi_2$,
and the lowest proton 2p-2h excitation $\pi\pi_1$. Their properties
are listed in Table~\ref{tab:62zn}.

\begin{figure}[thb]
\centering
\includegraphics[width=\columnwidth]{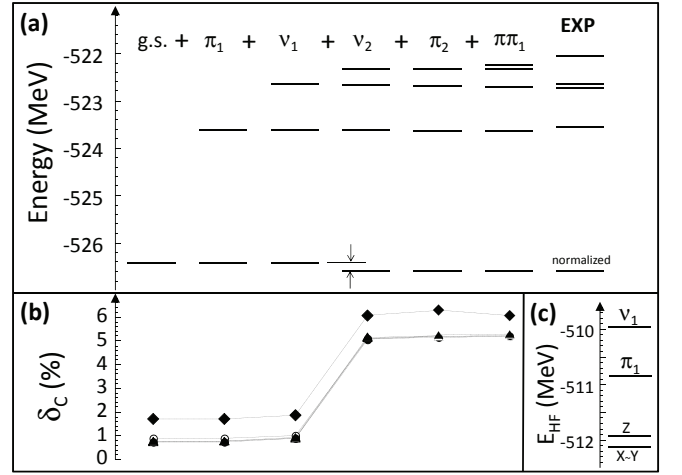}
\caption{(a) The low-lying $0^+$ states in $^{62}$Zn in function of
the number of configurations included in the NCCI calculations. (b)
Calculated ISB corrections versus the number of configurations taken
into account in the daughter nucleus. Different curves correspond to
different sets of configurations taken to calculate $0^+$ state in
$^{62}$Ga. (c) The HF energies of configurations included in the
calculation of $^{62}$Ga, see Table~\protect\ref{tab:62ga}. Further
details are given in the text.}
\label{fig:62ga}
\end{figure}

As discussed in Ref.~\cite{(Sat14d)}, the calculated spectrum of
$0^+$ states is in a very good agreement with the recent data
communicated by Leach {\it et al.\/}~\cite{(Lea13)}. As shown in
Fig.~\ref{fig:62ga}(a), the calculated total g.s.\ energy is stable
with increasing the number of reference configurations. Its value of
$-$526.595\,MeV  ($N=12$ harmonic oscillator shells were used)
underestimates the experiment by roughly 2\%.

\begin{table}[tbh]
\caption{Similar as in Table~\protect\ref{tab:Li}, but for $^{62}$Zn.
Here, the Slater determinants are labeled by neutron and proton
configurations described in the text. The last column shows
energies of the lowest $0^+$ states projected from a given
Slater determinant.}
\label{tab:62zn}
\renewcommand{\arraystretch}{1.3}
\centering\begin{tabular}{rrrrrrrcr}
\hline
$i$ & $|^{62}$Zn$;i\rangle$& $\Delta$E$_{\rm HF}$  &   $\beta_2$   &   $\gamma$    & $j_\nu$ & $j_\pi$ & $k$ &  $\Delta$E$_{\rm I=0}$ \\
\hline
 1  &    g.s.        &   $-$521.549          &      0.270    &  31$^\circ$   &    0.000 & 0.000    &    &    $-$526.405         \\
 2  &    $\pi_1$     &        1.433          &      0.286    &  20$^\circ$   &    0.005 & 0.152    & Y  &         2.036         \\
 3  &    $\nu_1$     &        3.347          &      0.255    &  40$^\circ$   &    0.689 & 0.318    & X  &         3.703         \\
 4  &    $\nu_2$     &        4.287          &      0.240    &  25$^\circ$   & $-$0.281 & $-$0.325 & Y  &         3.852         \\
 5  &    $\pi_2$     &        5.251          &      0.246     & 48$^\circ$   & $-$0.103 & $-$0.076 & X  &         5.672         \\
 6  &    $\pi\pi_1$  &        3.381          &      0.251     & 38$^\circ$   &    0.000 & 0.000    &    &         3.471         \\
\hline
\hline
\end{tabular}
\end{table}
\begin{table}[tbh]
\caption{Same as in Table~\ref{tab:62zn}, but for $^{62}$Ga.}
\label{tab:62ga}
\renewcommand{\arraystretch}{1.5}
\centering\begin{tabular}{rrrrrrrcr}
\hline
$i$ & $|^{62}$Ga$;i\rangle$\hspace*{-2mm}& $\Delta$E$_{\rm HF}$  &   $\beta_2$   &   $\gamma$    & $j_\nu$ & $j_\pi$ & $k$ &  $\Delta$E$_{\rm I=0}$ \\
\hline
 1  &    Y       &   $-$512.122          &      0.268    &     30$^\circ$           & $-$0.138 &    0.149 & Y   &     $-$516.930         \\
 2  &    X       &        0.007          &      0.268    &     30$^\circ$           &    0.180 & $-$0.170 & X   &       $-$0.001         \\
 3  &    Z       &        0.190          &      0.269    &     30$^\circ$           & $-$0.299 &    0.264 & Z   &          0.005         \\
 4  &    $\pi_1$ &        1.266          &      0.284    &     20$^\circ$           & $-$0.012 & $-$0.264 & X   &          2.175         \\
 5  &    $\nu_1$ &        1.977          &      0.255    &     35$^\circ$           & $-$0.440 & $-$0.351 & X   &          3.151         \\
\hline
\hline
\end{tabular}
\end{table}

In spite of the fact that the total binding energy is relatively
stable, the calculated ISB corrections to superallowed transition
$^{62}$Ga$\rightarrow ^{62}$Zn strongly depend on the details of the
calculation. This is illustrated in Fig.\ref{fig:62ga}(b), which
shows values of $\delta_{\rm C}$ in function of the number of
configurations taken for the NCCI calculations in the daughter
nucleus $^{62}$Zn. The four different curves correspond to different
model spaces taken for the NCCI calculation in the parent nucleus
$^{62}$Ga, see Table~\ref{tab:62ga} and Fig.~\ref{fig:62ga}(c).
In terms of Nilsson numbers, counted relatively to the $^{64}$Zn$_{32}$ even-even
core, the configurations X,Y,Z correspond to differently aligned
$\nu |312\, 3/2\rangle^{-1}\otimes \pi |312\, 3/2\rangle^{-1}$
two-hole states, $\pi_1$ denotes $\nu |312\, 3/2\rangle^{-1}\otimes \pi |312\, 5/2\rangle^{-1}$,
two hole state while $\nu_1$ is $\nu |321\, 1/2\rangle^{-1}\otimes \pi |312\, 3/2\rangle^{-1}$.
The three curves labeled with open dots, and open and filled triangles
correspond to states $0^+$ projected from the [X,Y], [X,Y,Z], and
[X,Y,Z,$\pi_1$] configurations, respectively. These curves
essentially overlap with each other, thus showing no influence of the
configuration-mixing (in this restricted model space) on the
structure of the $0^+$ state in the parent nucleus. Note, however,
that an extension of the model space by adding the lowest neutron p-h
excitation, [X,Y,Z,$\pi_1$,$\nu_1$], leads to an increase in
$\delta_{\rm C}$ of about 1\%. Note also, that all curves are
particularly sensitive to an admixture of the $\nu_2$ configuration
in the daughter nucleus. This admixture increases $\delta_{\rm C}$ by
almost 4\%. The analysis clearly shows that, within the present
implementation of the model, it is essentially impossible to match
the spaces of states used to calculate the parent and daughter
nuclei. The reasons are manifold. The lack of representability of the
$T=1,I=0$ states in the $N=Z$ nucleus within the conventional MF using
products of neutron and proton wave functions and difficulties in
constraining the time-odd part of the functional are two of them.
Difficulty of matching the model spaces in the parent and daughter
nuclei introduce here an artificial ISB effect. As a result, beyond a
simple mixing of orientations used in the result given in
Table~\ref{tab:delta1}, the NCCI approach cannot be used for
determining the ISB corrections to the transition
$^{62}$Ga$\rightarrow ^{62}$Zn.

\section{Summary and perspectives}
\label{sec:sum}

In this work, we introduced the NCCI model involving the isospin and
angular-momentum projections and subsequent mixing of states having
good angular momentum and properly treated Coulomb isospin mixing.
The model is capable of treating rigorously both the fundamental
(spherical, particle-number) as well as approximate (isospin) nuclear
symmetries. Its potentially unrestricted range of applicability and a
natural ability to treat the core-polarization effects resulting from
a subtle interplay between the long-range Coulomb force and
short-range hadronic nucleon-nucleon forces, which are treated on the
same footing, makes it an interesting alternative to the nuclear
shell-model.

The NCCI model employs states projected from low-lying
(multi)particle-(multi)hole deformed Slater determinants
(configurations) calculated self-consistently using Hartree-Fock
method. In the present realization, the
same SV Skyrme functional was used both to compute the
configurations and to mix the states. This restriction, however, can
be easily relaxed opening a room for various generalizations of the
model. In particular, one can attempt to correct an interaction used
at the mixing stage in order to improve a description of $T=0,I=1^+$
states in $^6$Li and $^{42}$Sc $N=Z$ nuclei.

We demonstrated that our NCCI formalism is capable of capturing many
features of the low-lying energy spectra in such diverse systems as
$^{8}$Li, $A$=38 isospin triplet nuclei, or $^{62}$Zn and $^{62}$Ga
nuclei. A reasonable agreement with experiment was obtained when
using a relatively small number of configurations, which supports our
claims that the model can indeed be applicable to medium heavy nuclei with
an affordable numerical cost. Our recent systematic study of
Gamow-Teller matrix elements in $T_z=1/2$ $sd$- and lower $pf$-shell
mirror nuclei performed in Ref.~\cite{(Kon15)}, see also
\cite{(Sat15)}, confirms that the model can incorporate in a
controlled way many important correlations into the nuclear wave
function.

Finally, we also calculated the new set of the ISB corrections to
superallowed $T=1,I=0^+\rightarrow T=1,I=0^+$ beta transitions. The
refined corrections are collected in Table~\ref{tab:delta1} for a
canonical set of precisely measured transitions and in
Table~\ref{tab:delta2} for transitions that were either unmeasured or
measured with the accuracy insufficient for the Standard Model tests.
These results are based on mixing the $I=0^+$ states projected from
the so-called $X$,$Y$, and $Z$ configurations corresponding to
different shape-current orientations in odd-odd nuclei.
Unfortunately, an attempt to perform more advanced calculation for
the transition $^{62}$Ga$\rightarrow ^{62}$Zn, which would take into
account more configurations, failed because of difficulties in
matching the model spaces in even-even and odd-odd nuclei.

\begin{acknowledgments}

This work was supported in part
by the Polish National Science Centre (NCN) under Contract Nos.\ 2012/07/B/ST2/03907 and 2014/15/N/ST2/03454,
by the THEXO JRA within the EU-FP7-IA project ENSAR (No.\ 262010),
by the ERANET-NuPNET grant SARFEN of the Polish National Centre for Research and Development (NCBiR),
and by the Academy of Finland and University of Jyv\"askyl\"a within the FIDIPRO programme.
We acknowledge the CSC-IT Center for Science Ltd., Finland, for the allocation of
computational resources.

\end{acknowledgments}

%

\end{document}